\documentclass[twocolumn,aps,pra,showpacs,groupedaddress]{revtex4}
\usepackage{epsfig,amssymb,amsmath,mathrsfs,color}
\usepackage[matrix,frame,arrow]{xypic}
\usepackage[matrix,frame,arrow]{xy}
\usepackage{amsmath}
\newcommand{\bra}[1]{\left\langle{#1}\right\vert}
\newcommand{\ket}[1]{\left\vert{#1}\right\rangle}
\newcommand{\qw}[1][-1]{\ar @{-} [0,#1]}
\newcommand{\qwx}[1][-1]{\ar @{-} [#1,0]}
\newcommand{\cw}[1][-1]{\ar @{=} [0,#1]}
\newcommand{\gate}[1]{*{\xy *+<.6em>{#1};p\save+LU;+RU **\dir{-}\restore\save+RU;+RD **\dir{-}\restore\save+RD;+LD **\dir{-}\restore\POS+LD;+LU **\dir{-}\endxy} \qw}
\newcommand{\measureD}[1]{*{\xy*+=+<.5em>{\vphantom{\rule{0em}{.1em}#1}}*\cir{r_l};p\save*!R{#1} \restore\save+UC;+UC-<.5em,0em>*!R{\hphantom{#1}}+L **\dir{-} \restore\save+DC;+DC-<.5em,0em>*!R{\hphantom{#1}}+L **\dir{-} \restore\POS+UC-<.5em,0em>*!R{\hphantom{#1}}+L;+DC-<.5em,0em>*!R{\hphantom{#1}}+L **\dir{-} \endxy} \qw}
\newcommand{\multimeasureD}[2]{*+<1em,.9em>{\hphantom{#2}}\save[0,0].[#1,0];p\save !C *{#2},p+LU+<0em,0em>;+RU+<-.8em,0em> **\dir{-}\restore\save +LD;+LU **\dir{-}\restore\save +LD;+RD-<.8em,0em> **\dir{-} \restore\save +RD+<0em,.8em>;+RU-<0em,.8em> **\dir{-} \restore \POS !UR*!UR{\cir<.9em>{r_d}};!DR*!DR{\cir<.9em>{d_l}}\restore \qw}
\newcommand{\control}{*!<0em,.025em>-=-{\bullet}}
\newcommand{\ctrl}[1]{\control \qwx[#1] \qw}
\newcommand{\multigate}[2]{*+<1em,.9em>{\hphantom{#2}} \qw \POS[0,0].[#1,0];p !C *{#2},p \save+LU;+RU **\dir{-}\restore\save+RU;+RD **\dir{-}\restore\save+RD;+LD **\dir{-}\restore\save+LD;+LU **\dir{-}\restore}
\newcommand{\ghost}[1]{*+<1em,.9em>{\hphantom{#1}} \qw}
\newcommand{\lstick}[1]{*!R!<.5em,0em>=<0em>{#1}}
\newcommand{\ustick}[1]{*!D!<0em,-.5em>=<0em>{#1}}
\newcommand{\Qcircuit}[1][0em]{\xymatrix @*[o] @*=<#1>}
\newcommand{\pureghost}[1]{*+<1em,.9em>{\hphantom{#1}}}
\newcommand{\multiprepareC}[2]{*+<1em,.9em>{\hphantom{#2}}\save[0,0].[#1,0];p\save !C
  *{#2},p+RU+<0em,0em>;+LU+<+.8em,0em> **\dir{-}\restore\save +RD;+RU **\dir{-}\restore\save
  +RD;+LD+<.8em,0em> **\dir{-} \restore\save +LD+<0em,.8em>;+LU-<0em,.8em> **\dir{-} \restore \POS
  !UL*!UL{\cir<.9em>{u_r}};!DL*!DL{\cir<.9em>{l_u}}\restore}
\newcommand{\puregate}[1]{*{\xy *+<.6em>{#1};p\save+LU;+RU **\dir{-}\restore\save+RU;+RD **\dir{-}\restore\save+RD;+LD **\dir{-}\restore\POS+LD;+LU **\dir{-}\endxy}}

\newcommand{\multipuregate}[2]{*+<1em,.9em>{\hphantom{#2}}  \POS[0,0].[#1,0];p !C *{#2},p \save+LU;+RU **\dir{-}\restore\save+RU;+RD **\dir{-}\restore\save+RD;+LD **\dir{-}\restore\save+LD;+LU **\dir{-}\restore}

\newtheorem{lemma}{Lemma} 
 
 \newtheorem{definition}{Definition}
\newtheorem{remark}{Remark}

\def\Lin{{\mathcal L}}
\def\d{\operatorname d}
\def\vec#1{\boldsymbol{#1}}
\def\qed{$\blacksquare$}
 
\def\>{\rangle}
\def\<{\langle}\def\kk{\>\! \>}\def\bb{\<\!\<}          
\def\Ket#1{|#1\>}
\def\Bra#1{\<#1|}

  \def\Tr{{\rm Tr}}
 
\def\ketbra#1#2{|#1\>\<#2|}
\def\group#1{\mathbb{#1}}
\def\hilb#1{\mathscr{#1}}
\def\defset#1{\mathsf{#1}}

\def\rank{\operatorname{Rnk}}
\def\lA{A}
\def\lB{B}
\def\lC{C}
\def\lD{D}
\def\lK{K}
\def\lL{L}

\begin{document}

\title{Cloning of a quantum measurement}
\author{Alessandro Bisio}\email{alessandro.bisio@unipv.it}
\affiliation{{\em QUIT Group}, Dipartimento di Fisica  ``A. Volta'' and INFN, via Bassi 6, 27100 Pavia, Italy}
\homepage{http://www.qubit.it}
\author{Giacomo Mauro D'Ariano}\email{dariano@unipv.it}
\affiliation{{\em QUIT Group}, Dipartimento di Fisica  ``A. Volta'' and INFN, via Bassi 6, 27100 Pavia, Italy}
\homepage{http://www.qubit.it}
\author{Paolo Perinotti}\email{paolo.perinotti@unipv.it}
\affiliation{{\em QUIT Group}, Dipartimento di Fisica  ``A. Volta'' and INFN, via Bassi 6, 27100 Pavia, Italy}
\homepage{http://www.qubit.it}
\author{Michal Sedl\'ak}\email{michal.sedlak@unipv.it}
\affiliation{{\em QUIT Group}, Dipartimento di Fisica  ``A. Volta'', via Bassi 6, 27100 Pavia, Italy}
\affiliation{Institute of Physics, Slovak Academy of Sciences, D\'ubravsk\'a cesta 9, 845 11 Bratislava, Slovakia}
\homepage{http://www.qubit.it}
\date{\today}
\begin{abstract}
  We analyze quantum algorithms for cloning of a quantum measurement.
  Our aim is to mimic two uses of a device performing an unknown von Neumann measurement with a single use of the device.
  When the unknown device has to be used before the bipartite state to be measured is available we talk about $1\rightarrow 2$ learning of
  the measurement, otherwise the task is called $1\rightarrow 2$ cloning of a measurement. We perform the optimization for both learning and cloning
  for arbitrary dimension of the Hilbert space. For $1\rightarrow 2$ cloning we also propose a simple quantum network that realizes the optimal
  strategy.
\end{abstract}
\maketitle

\section{Introduction}
Arbitrary processing of a classical information can be described by strings of bits, and can be performed by a fixed device for example a processor of any personal computer. As a consequence, we do not need to build new devices for different computations, but we just need to copy bit strings carrying the appropriate program. Situation dramatically changes, when the systems carrying the information are governed by quantum mechanics. Unknown states of quantum systems can not be copied perfectly \cite{wootzu} and the no-programming theorem \cite{noprog} prevents existence of universal quantum processors. This means that quantum programs can not be copied and that by using registers of qubits (two level quantum systems) one can not deterministically realize all quantum information processing functions with a fixed processor. So in contrast to classical devices, quantum ones cannot be replicated by just copying the program for them. Copying of quantum states was extensively investigated \cite{buzek1,werner1,cavesbroad1,superbroad1,scarani1}. On the other hand copying of quantum devices did not receive so much attention even though it is a fundamental and equally important quantum information processing task. Similarly to states, quantum transformations are often used in quantum key distribution schemes \cite{clonunit,pirandola1,bostrom1,lucamarini1} to encode bits, so analysis of possible attacks by cloning them are needed. Cloning of transformations was yet analyzed only for the case of unitary transformations \cite{clonunit}. In the present paper we investigate cloning of measurement devices, which can be seen as a cloning of certain measure-and-prepare transformations.
More precisely, when a measurement is an intermediate step of a quantum procedure its outcome can influence the following operations. This feed forward of the classical outcome can be conveniently described using a quantum system into which the outcome is encoded into perfectly distinguishable orthogonal states. In this sense a quantum measurement with only classical outcomes can be seen as a channel, which first measures the input system and based on the outcome prepares a state from a fixed orthogonal set.

The term cloning of observables has been used in Ref. \cite{paris1} 
referring to state cloning machines preserving the statistics of a class of observables.
In the present paper the objective is to actually mimic two uses of an unknown measurement device, while using it only once.
We would like to construct a replication strategy that would work for arbitrary  von Neumann measurement $\vec E$, even if it is provided as an unknown black box. The most natural operation of a replication strategy is based on modifying the bipartite state before $\vec E$ %the quantity $X$
is actually used. In this case we talk about \emph{$1 \rightarrow 2$ cloning} of a measurement device. The most general %network
representation of any cloning strategy is depicted below

%\emph{$1 \rightarrow 2$ cloning}: %The measurement device and the states we want to measure are available at the same time;
\begin{align}\label{sch:cloning}
  \begin{aligned}
    \Qcircuit @C=1em @R=1em {
      &\ustick{\lA}&\multigate{1}{\;\;\;}&\ustick{\lC}\qw&\measureD{\vec E}& \ustick{\lD} \cw & \pureghost{\;\;\;} \cw& \\
      &\ustick{\lB}&\ghost{\;\;\;} &\qw &\qw &\qw
      &\multimeasureD{-1}{\;\;\;}& }
  \end{aligned}
\end{align}
(the double wire carries the classical outcome of the measurement).

On the other hand, one might ask %whether %at least approximatively
how well the task can be accomplished when we use the measurement, before we have an access to the bipartite state of interest. We denote this scenario as \emph{$1 \rightarrow 2$ learning}, and any learning strategy can be depicted as follows
%and a network of any learning strategy is as follows.

%\emph{$1 \rightarrow 2$ learning}: we can use the measurement device
%only once today and we want to replicate the same observable on two
%systems that will be provided tomorrow
\begin{align}\label{sch:learning}
  \begin{aligned}
    \Qcircuit @C=0.7em @R=1em { & &
      & \ustick{\lA}& \ghost{\;\;\;}\\
      &&& \ustick{\lB}& \ghost{\;\;\;}\\
      \multiprepareC{1}{\;\;\;} &\ustick{\lC} \qw & \measureD{\vec E} & \ustick{\lD} \cw &\pureghost{\;\;\;} \cw\\
      \pureghost{\;\;\;}& \qw& \qw&\qw & \multimeasureD{-3}{\;\;\;}\\
    }
  \end{aligned} \quad .
\end{align}
In the present paper we will analyze only the above two scenarios, even
though one can think of more general versions of the problem, where the $M$ replicas have to be produced out of $N$ uses of a measurement device.
For example $N\rightarrow 1$ learning was analyzed in Ref. \cite{learnobs}.
From comparison of Eqs. (\ref{sch:cloning}) and (\ref{sch:learning}) one can see that learning is a particular instance of cloning in which the first step is restricted. That being so, it is clear that the performance of the optimal learning cannot be better than the performance of the optimal cloning.

The paper is organized as follows.
In Sec. \ref{schemform} we expose the formulation of the optimal learning and cloning in mathematical terms.
In Sec \ref{prelim} we review the framework of quantum combs that is
used as main tool throughout the paper.
In Sec. \ref{sec:obssymm-repl-netw} the problem is simplified exploiting all the symmetries that can be useful. Sections
\ref{sec:clon}, \ref{sec:learning} are devoted to derivation of optimal cloning and learning, respectively.
The paper is closed by concluding remarks in Sec. \ref{conc}.

\section{Mathematical formulation of the problem}\label{schemform}

Let us now formulate the problem mathematically.
First of all, we should be able to evaluate the performance of the chosen replication
strategy $\vec {\mathcal R}$. Hence, we need a quantity that expresses
the closeness of a replicated measurement to a desired bipartite von
Neumann measurement.
In the following Lemma we introduce a function $\mathscr{F}(\vec P,\vec Q)$ that
 quantifies the closeness of a POVM $\vec Q$ to a von Neumann POVM $\vec P$.
Throughout the paper we shall use the bold face notation for objects that are composed from several elements. For example $\vec P\equiv \{P_i\}_{i=1}^d$ denotes the POVM with elements $P_i$ and ${\vec I} \equiv \{ I\}$ is the single outcome POVM.
%The interpretation of $\mathscr{F}(\vec P,\vec Q)$ as a "fidelity" between $\vec P$ and $\vec Q$ is provided by the following Lemma.
\begin{lemma}[Fidelity criterion for POVM]
  Let $\Sigma := \{ 1,\dots,d \}$ be a finite set of events and $\vec P\subseteq \mathcal{L}(\hilb{H})$ and $\vec Q\subseteq\mathcal{L}(\hilb{H}) $ be two POVM's, such that one of them is a von Neumann measurement.
  Consider now the quantity
\begin{equation}
   \mathscr{F}(\vec P,\vec Q) := \frac1d \sum_{i=1}^d \Tr[P_iQ_i].
   \label{eq:fidelity}
\end{equation}
  Then $\mathscr{F}=1
  \Leftrightarrow P_i = Q_i\  \forall i$ and $\mathscr{F}\leq 1$.
\end{lemma}

\begin{Proof}
  %If
Without loss of generality we can assume that $\vec P$ is a von Neumann measurement and that we have $P_i=
  \ketbra{i}{i}$ where $\ket{i}$ is an orthonormal basis of
  $\hilb{H}$.  Then for $Q_i = P_i= \ketbra{i}{i}$ we have
  \begin{align}
    \mathscr{F}= \frac1d \sum_{i=1}^d \Tr[P_iQ_i] = \frac1d
    \sum_{i=1}^d \Tr[\ketbra{i}{i}] = 1.
  \end{align}
  On the other hand if $\mathscr{F}=1$ we have
  \begin{align}
    d=& \sum_{i=1}^d \Tr[P_iQ_i] = \sum_{i=1}^d \bra{i}Q_i\ket{i} = \nonumber\\
    &\sum_{i,j=1}^d \bra{i}Q_j\ket{i} - \sum_{i\neq j} \bra{i}Q_j\ket{i}= \nonumber \\
    &\Tr\left[ \sum_{j=1}^d Q_j \right] - \sum_{i\neq j} \bra{i}Q_j\ket{i}=\nonumber\\
    &d - \sum_{i\neq j} \bra{i}Q_j\ket{i},
  \end{align}
  which %finally
  implies $\sum_{i\neq j} \bra{i}Q_j\ket{i} = 0$. Since
  $Q_j \geq 0$, we must have $\bra{i}Q_j\ket{i}=0$ for all $i \neq j$,
  and consequently $Q_j = \alpha_j \ketbra{j}{j}$ with $\alpha_j \geq 0$.
  Finally the condition $\sum_{j=1}^d \alpha_j \ketbra{j}{j} =\sum_{j=1}^d Q_j= I$
  implies $\alpha_j = 1$ and thus $Q_j = P_j$. Proving that $\mathscr{F}\leq 1$ is easy. Since $Q_i$ is an element of a POVM we have $\bra{i}Q_i\ket{i}\leq 1$ and consequently $\mathscr{F}=\frac1d \sum_{i=1}^d \bra{i}Q_i\ket{i}\leq 1$.
\qed
\end{Proof}

Since we assume that the unknown measurement $\vec E$ to be replicated
is a von Neumann POVM, we can write it in the following form
\begin{align}
  E_i = \ketbra{\phi_i}{\phi_i}
\end{align}
where $\{ \ket{\phi_i} \}_{i=1}^d$ is an orthonormal basis of the Hilbert space
$\mathcal{H}$.  All the POVM's of this kind can be generated by
rotating a reference POVM $\{\ketbra{i}{i}\}_{i=1}^d$ by
elements of the group of unitary transformations $\group{SU}(d)$ as
follows
\begin{align}
  E_i^{(U)} = U\ketbra{i}{i}U^\dagger \qquad U \in \group{SU}(d).
\end{align}

Let us denote the bipartite POVM replicated by the strategy $\boldsymbol{\mathcal{R}}$ as $\vec G^{(U)} \equiv \vec G(\boldsymbol{\mathcal{R}},{\vec E}^{(U)})$.
Our task is to find such replicating strategy $\boldsymbol{\mathcal{R}}$ that the elements $G^{(U)}_{ij}$ are as close as possible to $E_{i}^{(U)} \otimes E_{j}^{(U)}$.
Assuming that the unknown POVM $\vec E^{(U)}$ is randomly drawn
according to the Haar distribution, we choose the quantity:
\begin{align}\label{eq:obsfigmer}
  F[\boldsymbol{\mathcal{R}}] &:= \int \d U \; \mathscr{F}( \vec G^{(U)},\vec E^{(U)}\otimes \vec E^{(U)})= \\
&=\frac{1}{d^2}\sum_{i,j=1}^d \int \d U \; \Tr[G^{(U)}_{ij} (E^{(U)}_i\otimes E^{(U)}_j)]
\nonumber
\end{align}
as a figure of merit for the replicating strategy. Hence, after choosing
one of the two considered scenarios ($1
\rightarrow 2$ cloning or learning) the goal is to find a strategy $\boldsymbol{\mathcal{R}}$, that maximizes $F[\boldsymbol{\mathcal{R}}]$.

\section{Preliminary concepts}\label{prelim}
In this section we introduce the necessary notation and review the general theory of
\emph{Quantum Networks}, as developed in \cite{architecture,comblong}.
Let us first recall the Choi-Jamio\l kowsky isomorphism. It is an isomorphism connecting any quantum operation (i.e. completely positive map)
 $\mathcal{M}:\mathcal{B}(\hilb{H}_{in}) \mapsto \mathcal{B}(\hilb{H}_{out})   $
to a positive operator $M \in \mathcal{B}(\hilb{H}_{out} \otimes \hilb{H}_{in})$
defined as follows:
\begin{align}
  M := \mathcal{M} \otimes \mathcal{I} (\ketbra{\omega}{\omega})
\end{align}
where $\mathcal{I}$ is the identical map on $\mathcal{B}(\hilb{H}_{in})$,
$\ket{\omega} := \sum_n \ket{n}\ket{n} \in \hilb{H}_{in} \otimes \hilb{H}_{in}$
and we fixed an orthonormal basis $\{ \ket{n} \}$ on $\hilb{H}_{in}$.
The action of $\mathcal{M}$ on a given input state $\rho$
can be expressed in terms  of  $M$ as:
\begin{align}
  \mathcal{M}(\rho) = \Tr_{in}[M(I \otimes \rho^T)]
\end{align}
where $\Tr_{in}$ denotes the partial trace over $\hilb{H}_{in}$ and
the superscript $T$ marks the transposition with respect to the basis $\{ \ket{n} \}$.

%We now review the general theory of \emph{Quantum Networks}, as developed in \cite{architecture,comblong}.

Under %By
the term Quantum Network we mean a network of quantum devices part of whose inputs and outputs are %mutually
connected, while the remaining
%pairs of outputs and inputs
ones are forming open slots of the network into which %variable
sub-circuits can be later inserted.
  A network with $(N-1)$ open slots has $N$ input and $N$ output
systems, that we label by even numbers from $0$ to $2N-2$ and by odd
numbers from $1$ to $2N-1$, respectively. Each network can be visualized as in Eq.  (\ref{Eq:comb}),
\begin{equation}
  \Qcircuit @C=1em @R=1em {
    \ustick{0}&\multigate{1}{C_1}&\ustick{1}\qw&&\ustick{2}\qw&\multigate{1}{C_2}&\ustick{3}\qw&\ghost{\dots}&\ustick{2N-2}\qw&\multigate{1}{C_N}&\ustick{2N-1}\qw\\
    &\pureghost{C_1}&\qw&\qw&\qw&\ghost{C_2}&\qw&\cdots&&\ghost{C_N}&}
  \label{Eq:comb}
\end{equation}
where the wires represent the connections of output systems to next inputs.
This flow of quantum systems induces a {\em causal
  order} among the wires , according to which the input system $m$ cannot influence
the output system $n$ if $m>n$.

A Quantum Network $\mathcal{R}$ can be represented in terms of its
Choi-Jamio\l kowsky operator $R$, called {\em quantum comb},
which is a positive operator acting on the Hilbert
space $\hilb H_{out} \otimes \hilb{H}_{in}$ where $\hilb{H}_{out} :=
\bigotimes_{j=0}^{N-1} \hilb{H}_{2j+1}$, $\hilb{H}_{in} := \bigotimes_{j =
  0}^{N-1} \hilb{H}_{2j}$, and $\hilb{H}_n$ being the Hilbert space of the $n$-th
system. For a deterministic quantum network (i.e. a network of quantum
channels) the causal structure implies the following
normalization condition
\begin{equation}\label{recnorm}
\Tr_{2k-1} [ R^{(k)}] = I_{2k-2} \otimes R^{(k-1)} \qquad k=1, \dots, N~
\end{equation}
where $R^{(N)}=R$, $R^{(0)} =1$, $R^{(k)}
\in\Lin (\bigotimes_{n=0}^{2k-1} \hilb{H}_n)$, $\Tr_{2k-1}$ denotes the partial trace on $\hilb{H}_{2k-1}$ and $I_{2k-2}$ is an identity operator on $\hilb{H}_{2k-2}$.

We can also consider probabilistic quantum networks (i.e. networks
of quantum operations), whose
Choi-Jamio\l kowsky operators
must satisfy
\begin{align}
  \label{eq:probcomb}
  0 \leq R \leq S
\end{align}
where $S$ is the Choi-Jamio\l kowsky
operator of a deterministic network.

We call \emph{generalized instrument}
a set of probabilistic quantum networks
$\vec {\mathcal{R}} := \{ \mathcal{R}_i\}$
such that the set ${\bf R} := \{ R_i \}$
of the corresponding Choi operators satisfies
\begin{align}
  \label{geninst}
  \sum_i R_i = R_\Omega
\end{align}
where $R_\Omega$ is the Choi operator of a deterministic network.

Two quantum networks $\mathcal R_1$
and $\mathcal R_2$ can be connected by linking some outputs of
$\mathcal R_1$ ($\mathcal R_2$) with
inputs of $\mathcal R_2$ ($\mathcal R_1$), thus forming a new network $\mathcal R_3 :=\mathcal  R_1 * \mathcal R_2$. We adopt
the convention that the wired to be connected are identified by the
same label. The connection of the two quantum networks is mathematically represented by the {\em link
  product} of the corresponding Choi operators $R_1$ and $R_2$, which is
defined as
\begin{equation}\label{link-prod-def}
  R_1 * R_2 =\Tr_{\hilb{K}}[R_1^{\theta_{\hilb{K}}} R_2],
\end{equation}
$\theta_{\hilb{K}}$ denoting partial transposition over the Hilbert space
$\hilb{K}$ of the connected systems (recall that we identify the Hilbert spaces of connected systems with the same
labels).

%%%%%%%%%%%%%%%%%%%%%%%%%%%%%
As we pointed out in the introduction,  the classical outcome of the
inserted measurement can influence the next operation of the network.
In order to take the feed forward of the classical outcome into account
it is convenient to describe the measurement device to be replicated
as a measure-and-prepare quantum channel
%$\mathcal{E}^{(U)}:\mathcal{L}(\hilb{H})\longrightarrow \mathcal{L}(\hilb{H})$
\begin{align}
  \mathcal{E}^{(U)}(\rho) = \sum_{i=1}^d\Tr[E_i^{(U)}\rho]\ketbra{i}{i},
\end{align}
 which measures the POVM ${\vec E}^{(U)}$ on the input state and
in the case of outcome $i$ prepares the state $\ket{i}$ from a fixed orthonormal basis on the
output of the channel.
Within this framework the classical outcome is encoded into a
quantum system by preparing it into a state from a set of orthogonal
states.
The Choi-Jamio\l kowski representation of the channel $\mathcal{E}^{(U)}$
is the following
\begin{align}\label{eq:obsdepolachan}
  {E}^{(U)} = \sum_{i=1}^d \ketbra{i}{i} \otimes {E_i^{(U)}}^T = \sum_{i=1}^d \ketbra{i}{i}\otimes U^* \ketbra{i}{i} U^T,
\end{align}
where $X^T$ denotes the transpose of $X$ in the basis $\{\ket
i\}_{i=1}^d$.

Since we want the replicating network $\boldsymbol{\mathcal{R}}$ to behave as two
copies of the POVM ${\vec E}^{(U)}$ upon insertion of a single use
of $\mathcal{E}^{(U)}$, we have that $\boldsymbol{\mathcal{R}}$ is actually a
generalized instrument $\vec R \equiv\{ R_{ij} \}_{i,j=1}^d$ where $i,j$ is
the couple of outcomes of the two replicated measurements.
The normalization of the generalized instrument $\vec R = \{ R_{ij} \in \mathcal{L}( \hilb{H}_{\lA}\otimes \hilb{H}_{\lB}\otimes\hilb{H}_{\lC}\otimes\hilb{H}_{\lD}) \}$ has to obey the following equations:

\noindent\emph{$1\rightarrow 2$ cloning}
\begin{align}
   \sum_{i,j} R_{ij} = R_\Omega = I_\lD \otimes S_{\lA \lB \lC} \quad \Tr_\lC[S] = I_{\lA \lB},
\label{eq:normclon}
\end{align}
\noindent \emph{$1\rightarrow 2$ learning}
\begin{align}
   \sum_{i,j} R_{ij} = R_\Omega = I_{\lA \lB \lD} \otimes \rho_{\lC} \quad \Tr[\rho] = 1,
\label{eq:normlearn}
\end{align}
where the capital subscripts denote the Hilbert spaces on which the operators act and we use the labeling %of the Hilbert spaces
introduces in the Eqs. (\ref{sch:cloning}), (\ref{sch:learning}).

The replicated POVM is then equal to
\begin{align}\label{eq:obsfinalpovm}
  G^{(U)}_{ij} &= {[ R_{ij}*{E}_{\lC \lD}^{(U)}]}^T  \\
%\\  \nonumber &R_{ij} \in \mathcal{L}( \hilb{H}_{\lA}\otimes \hilb{H}_{\lB}\otimes\hilb{H}_{\lC}\otimes\hilb{H}_{\lD}).
%  &=[\bra{kk}U^\dagger \otimes I R_{ij} U\otimes I\ket{kk}_{\lC \lD}]^T ,
  &=\left[\sum_k \bra{k}_{\lC}\bra{k}_{\lD} (U^\dagger \otimes I) R_{ij} (U\otimes I) \ket{k}_{\lC} \ket{k}_{\lD}\right]^T .\nonumber
\end{align}

\section{Symmetries of the replicating network}\label{sec:obssymm-repl-netw}
In this section we utilize the symmetries of the figure of merit
(\ref{eq:obsfigmer}) to simplify the optimization problem. These considerations apply both to cloning and learning of a measurement device. The first
simplification relies on the fact that some wires of the network carry
only classical information, representing the outcome of the
measurement. The classical information encoded in the choice of a state from basis $\{\ket{i}\}$ can be read without disturbance by the measure and prepare channel $\mathcal{M}$ with Choi-Jamiolkowski operator $M\equiv E^{(I)}$. Thus, inserting channel $\mathcal{M}$ between the use of a measurement device $E^{(U)}$ and the network $\vec {\mathcal R}$ will not change the operation of the scheme, i.e.
\begin{align} \label{diagcirc}
\begin{aligned}
    \Qcircuit @C=0.5em @R=1em {
      &&\multigate{1}{\;\;\;} &\measureD{\vec E}& \puregate{M} \cw &\pureghost{\;\;\;} \cw& \\
      &&\ghost{\;\;\;} &\qw &\qw
      &\multigate{-1}{\;\;\;}& }
  \end{aligned}  
=
\begin{aligned}
    \Qcircuit @C=0.5em @R=1em {
      &\multigate{1}{\;\;\;} &\measureD{\vec E} & \pureghost{\;\;\;} \cw& \\
      &\ghost{\;\;\;} &\qw
      &\multigate{-1}{\;\;\;}& }
  \end{aligned} 
\end{align}
As a consequence we have the following lemma.

\begin{lemma}[Restriction to diagonal network]
  The optimal generalized instrument $\vec {\mathcal R}$, $\sum_{i,j}
  R_{ij} = R_{\Omega}$ maximizing Eq.
  (\ref{eq:obsfigmer}) can be chosen to satisfy:
\begin{align}\label{eq:obsdiagcomb}
  R_{ij} = \sum_{k}  R'_{ij,k} \otimes \ketbra{k}{k}_{\lD},
\end{align}
where $0\le R'_{ij,k} \in \mathcal{L}( \hilb{H}_{\lA}\otimes \hilb{H}_{\lB}\otimes\hilb{H}_{\lC})$.
\end{lemma}
\begin{Proof}
Let $S_{ij}$ be the Choi representation of a generalized instrument corresponding to a quantum network $\vec {\mathcal{S}}$.
  Let us define network $\vec {\mathcal{R}}$ as
  \begin{align}
    {R}_{ij} := \sum_{k} \bra{k} S_{ij}\ket{k} \otimes \ketbra{k}{k},
  \end{align}
  which can be seen as ${R}_{ij} = S_{ij}*M= S_{ij}*E^{(I)}$ (see Eq. (\ref{eq:obsdepolachan})) with the link performed on system $\lD$ carrying the classical information.
  We can easily prove that $\vec{ \mathcal{R}}$ is a generalized instrument. Indeed we have
  \begin{align}
    \sum_{i,j} {R}_{ij} =&\sum_{i,j} S_{ij}*E^{(I)}=S_{\Omega}* E^{(I)},\label{decchoi}
  \end{align}
  where the link is performed only on the space $\hilb{H}_{\lD}$.
  The operator in Eq.~\eqref{decchoi} is the Choi-Jamio\l kowski operator of a
  deterministic quantum network satisfying the same normalization
  conditions as $S_\Omega$.  Since $M*E^{(U)}=E^{(U)}$ we show that $\vec{ \mathcal{S}}$ %$\vec S$
  and $\vec{ \mathcal{R}}$ %  $\vec {R}$
  produce the same replicated POVM $G_{ij}^{(U)}$
  when linked with the single use of $E^{(U)}$, as follows
  \begin{align}
    (G^{(U)}_{ij})^T &= S_{ij}* {E}_{\lC \lD}^{(U)}=S_{ij}*M*{E}_{\lC \lD}^{(U)}  =  R_{ij}* {E}_{\lC \lD}^{(U)} \nonumber \\
    & =\sum_{k} ( \bra{k}_{\lD}\bra{k}_{\lC} U^\dagger)
      S_{ij} ( \ket{k}_{\lD} U \ket{k}_{\lC}),
      \label{eq:reppovm}
  \end{align}
where the explicit form of the star product will be used later.
The thesis then holds with $R'_{ij,k}:=\Bra{k}S_{ij}\Ket{k}$.\qed
\end{Proof}

The restriction to diagonal networks allows us  to simplify the
figure of merit (Eq.  (\ref{eq:obsfigmer})) as follows

\begin{eqnarray}\label{eq:obsfigmer2}
  F[\boldsymbol{\mathcal{R}}] &:=&    \int \d U \mathscr{F}( \vec{G}^{(U)}, \vec {E}^{(U)}\otimes \vec {E}^{(U)})  = \nonumber \\
  &=&  \frac{1}{d^2}\int \d U \sum_{i,j,k}
  \Tr\left[ R'_{ij,k}\times \right.  \\
 & &\left. \times {U^*}^{\otimes 2}\otimes{U}\ket{ijk}\bra{ijk}
  {U^T}^{\otimes 2}\otimes{U^\dagger}\right], \nonumber
\end{eqnarray}
where $\ket{ijk}\equiv \ket{i}_{\lA}\ket{j}_\lB \ket{k}_\lC$ and we applied Eq. (\ref{eq:obsdiagcomb}).

Since the performance of the scheme is evaluated as an average over all possible "orientations" of the replicated measurement device, there exists a symmetrization procedure that can make any strategy covariant (i.e. having property from Eq (\ref{covcirc})), without affecting the figure of merit.

\begin{align} \label{covcirc}
  \begin{aligned}
    \Qcircuit @C=0.5em @R=1em {
      &&\gate{U}&\multigate{1}{\;\;\;} &\measureD{\vec E} & \pureghost{\;\;\;} \cw& \\
      &&\gate{U}&\ghost{\;\;\;} &\qw
      &\multigate{-1}{\;\;\;}& }
  \end{aligned}
=
  \begin{aligned}
    \Qcircuit @C=0.5em @R=1em {
      &&\multigate{1}{\;\;\;}&\gate{U}&\measureD{\vec E}& \pureghost{\;\;\;} \cw& \\
      &&\ghost{\;\;\;} &\qw &\qw
      &\multigate{-1}{\;\;\;}& }
  \end{aligned}
\end{align}
This translates into  mathematical terms as follows.
\begin{lemma}[Restriction to covariant networks]
  The operators $R'_{ij,k}$ that maximize Eq.
  (\ref{eq:obsfigmer2}) can be chosen to satisfy the commutation
  relation
  \begin{align}\label{eq:obscommutr}
    [R'_{ij,k}, U^*_{\lA}\otimes U^*_{\lB}\otimes
    U_{\lC}]=0.
  \end{align}
\end{lemma}
\begin{Proof}
  Suppose that the generalized instrument corresponding to $S'_{ij,k}$ is optimal. Then one can easily
  check that also the instrument $R'_{ij,k}$
  defined as follows
  \begin{equation}
    R'_{ij,k}:=\int \d U ({U^*}^{\otimes 2}\otimes{U})S'_{ij,k}({U^T}^{\otimes 2}\otimes{U^\dagger})
  \end{equation}
  is suitably normalized and satisfies $[R'_{ij,k}, U^*\otimes U^*\otimes U]=0$.
  Generalized instrument $\vec {\mathcal R}$ corresponds to a strategy where random unitary $U^\dagger$, $U^\dagger$, $U$ is applied before and after the original strategy $\vec {\mathcal S}$ to systems $\lA$, $\lB$, $\lC$, respectively.
  From the integration in Eq. (\ref{eq:obsfigmer2}) it is obvious that the value of $F$ for the above choice of $R'_{ij,k}$ is the same as for
  $S'_{ij,k}$.\qed
\end{Proof}

The commutation relation (\ref{eq:obscommutr}) allows us to rewrite
the figure of merit as
\begin{align}\label{eq:obsfigmer3}
  F[\boldsymbol{\mathcal{R}}] = \frac{1}{d^2}\sum_{i,j,k}
  \bra{ijk}R'_{ij,k} \ket{ijk}_{\lA \lB \lC}.
\end{align}
Another symmetry we can utilize is related to a
simultaneous relabeling of the outcomes of the inserted and produced measurements. We shall denote by $\sigma$ the
element of $\group{S}_d$, the group of permutations of $d$ elements,
and by $T_\sigma$ the linear operator that permutes the elements of basis
$\{\ket{i}\}$ according to this permutation, in formula
$T_\sigma\ket{i}=\ket{\sigma(i)}$. Let us note that the complex conjugation and transposition are defined with respect to the basis $\{\ket{i}\}$, so $T_\sigma=T^*_\sigma$.

\begin{lemma}[Relabeling symmetry]
  Without loss of generality we can assume that the operators
  $R'_{ij,k}$ that maximize Eq. (\ref{eq:obsfigmer2})
  satisfy the relation
  \begin{align}\label{eq:obsperminvproperty}
    R'_{ij,k} = R'_{\sigma(ij,k)}.
  \end{align}
  \label{lem:relabsym}
  where we shortened
  $\sigma(ij,k):=(\sigma(i)\; \sigma(j),\sigma(k))$.

\end{lemma}

\begin{Proof}
Suppose that network $\vec{\mathcal{S}}$ characterized by operators $S_{ij}$ is optimal and
  satisfies both conditions \eqref{eq:obsdiagcomb} and
  \eqref{eq:obscommutr}. Let us then define
  \begin{eqnarray}
    R'_{ij,k}&:=&\frac{1}{d!}
    \sum_{\sigma\in \group{S}_d} ({T_\sigma^\dagger}^{\otimes 2}\otimes {T_\sigma}^\dagger) S'_{\sigma(ij,k)}({T_\sigma}^{\otimes 2}\otimes {T_\sigma})\nonumber\\
    &=&\frac{1}{d!} \sum_{\sigma\in \group{S}_d} S'_{\sigma(ij,k)},
    \label{eq:obscompatsym}
  \end{eqnarray}
  where the last identity in (\ref{eq:obscompatsym}) follows from the
  commutation relation (\ref{eq:obscommutr}) with $ U = T_\sigma$.
  The operators $R'_{ij,k}$ correspond to a valid quantum network $\vec{\mathcal{R}}$, because $\vec{\mathcal{R}}$ is a
  convex combination of networks $\vec{ \mathcal{Z}}^\sigma$ defined by
  Eq.~\eqref{eq:obsdiagcomb} with $Z'^\sigma_{ij,k}=S'_{\sigma(ij,k)}$.
  Quantum network $\vec{\mathcal{R}}$ operationally
  corresponds to relabeling of the outcomes of the inserted and
  replicated measurements by permutation $\sigma$. The figure of merit
  for $\vec{\mathcal{R}}$ is
  \begin{align}
    &F[\boldsymbol{\mathcal{R}}]=\frac{1}{d^2}\sum_{i,j,k}
    \bra{ijk} R'_{ij,k}
    \ket{ijk} = \nonumber\\%
    &\frac1{d^2 d!} \sum_{\sigma\in \group{S}_d} \sum_{i,j,k}
    \bra{\sigma(ijk)}
    S'_{\sigma(ij,k)}
    \ket{\sigma(ijk)}=F[\boldsymbol{\mathcal{S}}].
  \end{align}
  It is easy to prove that $R'_{ij,k}$ satisfy Eq.
  (\ref{eq:obsperminvproperty}).\qed
\end{Proof}

\begin{remark}
  The properties (\ref{eq:obsdiagcomb}), (\ref{eq:obscommutr}) and
  (\ref{eq:obsperminvproperty}) induce the following structure of the
  replicated POVM's:
  \begin{align}
    G^{(U)}_{\sigma(ij)} = {(U T_\sigma)}^{\otimes 2}
    G^{(I)}_{ij}{(T^\dagger_\sigma U^\dagger)}^{\otimes 2}
  \end{align}
\end{remark}
The advantage of using the relabeling symmetry is the reduction of the
number of independent parameters of the quantum generalized
instrument. Let us define the equivalence relation between strings
$ijk$ and $i'j'k'$ as
\begin{equation}
  ijk\sim\ i'j'k'\quad\Leftrightarrow\quad ijk=\sigma(i'j'k'),
\end{equation}
for some permutation $\sigma$. Thanks to Eq.
(\ref{eq:obsperminvproperty}) there are only as many independent
$R'_{ij,k}$ as there are equivalence classes among sequences
$(ij,k)$. There are four or five equivalence classes depending on the
dimension $d$ being two or greater than two, respectively. We denote
the set of these equivalence classes  by $\defset L:=\{xxx, xxy, xyx, xyy, xyz\}$.

Based on lemma \ref{lem:relabsym} we can write the optimal
generalized instrument as follows
\begin{equation}
 R_{ab,c}:=R'_{ij,k}=R'_{\sigma(ij,k)},
\end{equation}
where $(ab,c)$ is a string of indices that
represents one equivalence class from $\defset L$.

The figure of merit can finally be written as follows
\begin{equation}
  F[\boldsymbol{\mathcal{R}}]=\frac1{d^2}\sum_{(ab,c)\in\defset L} n(ab,c)\<R_{ab,c}\>,
  \label{eq:finfigm}
\end{equation}
where $n(ab,c)$ is the cardinality of the equivalence class
denoted by $(ab,c)$, and $\<R_{ab,c}\>=\bra {ijk}R'_{ij,k}\ket{ijk}$ for any string $ijk$ in the equivalence class denoted
by $(ab,c)$. As a consequence of Schur's lemmas, Eq.~\eqref{eq:obscommutr} implies the following structure
for the operators $R_{ab,c}$
\begin{equation}\label{eq:irrdecomp}
  R_{ab,c}=\bigoplus_\nu P^\nu \otimes r^\nu_{ab,c},
\end{equation}
where $\nu$ labels the irreducible representations in the
Clebsch-Gordan series of $U^*_{\lA}\otimes U^*_{\lB}\otimes U_{\lC}$, and $P^\nu$ acts as the identity on
the invariant subspaces of the representations $\nu$, while
$r^\nu_{ab,c}$ acts on the multiplicity space of the same
representation.

Depending on the dimension $d=2$ or $d>2$ we have two different decompositions. In the former case, we have
\begin{align}\label{eq:obsdecompor21dim2}
  R_{ab,c} = P^\alpha\otimes r^\alpha_{ab,c} +
  P^\beta r_{ab,c}^\beta,
  %  \label{eq:rxyinirreps}
\end{align}
where $r_{ab,c}^\alpha$ is a positive 2$\times$2 matrix,
while $r_{ab,c}^\beta$ is a non-negative real number. The
projections $P^\xi$ on the invariant spaces of the representation %$U^*\otimes U\otimes U$
$U^*\otimes U^*\otimes U$ are the following
\begin{align}
  &P^\alpha\otimes\ketbra ij=\sum_{m=1}^d|\Psi^i_m\kk \bb \Psi^j_m|,\quad i,j\in\{+,-\}\nonumber\\
  &P^\beta=I\otimes P^+-P^\alpha\otimes\ketbra++,
\end{align}
where $\Psi^\pm_m=(\ket{\omega} \Ket m\pm \Ket m\ket{\omega})/[2(d\pm1)]^\frac12$,
and $P^+$, $P^-$, are the projections onto the symmetric and
antisymmetric subspace, respectively. When $d>2$, on the other hand, we have
\begin{align}\label{eq:obsdecompor21}
  R_{ab,c} = P^\alpha \otimes r_{ab,c}^\alpha+
  P^\beta r_{ab,c}^\beta + P^\gamma r_{ab,c}^\gamma,
\end{align}
where $r_{ab,c}^\alpha$ is a positive 2$\times$2 matrix,
while $r_{ab,c}^\beta$ and $r_{ab,c}^\gamma$ are
non-negative real numbers. The projections $P^\xi$ on the invariant
subspaces %$U^*\otimes U^*\otimes U$)
are the following
\begin{align}
  &P^\alpha\otimes\ketbra ab=\sum_{m=1}^d|\Psi^a_m\kk \bb \Psi^b_m|,\quad a,b\in\{+,-\}\nonumber\\
  &P^\beta=I\otimes P^+-P^\alpha\otimes\ketbra++,\nonumber\\
  &P^\gamma=I\otimes P^--P^\alpha\otimes\ketbra--.
\end{align}

The last symmetry we are going to introduce relies on the
possibility to exchange the inputs (Hilbert spaces $\hilb{H}_\lA$
and $\hilb{H}_\lB$) of the two replicated measurements with simultaneously exchanging their measurement outcomes, while the figure of merit is left unchanged.

\begin{lemma}\label{lem:obs12perminv2}
  The operators $R_{ab,c}$ in Eq. (\ref{eq:irrdecomp}) can be chosen to satisfy
\begin{align}\label{eq:obs12permsymmetry}
  R_{ab,c} = \mathsf{S}R_{ba,c}\mathsf{S} \quad \forall (ab,c)\in \defset L
\end{align}
where $\mathsf{S}$ is the swap operator $\mathsf{S}\ket{k}_\lA \ket{j}_\lB = \ket{j}_\lA \ket{k}_\lB$.
\end{lemma}
\begin{Proof}
  The proof can be done by the following averaging argument.  Let us
  define $\overline{R}_{ij,k}:=\frac12 (R'_{ij,k} + \mathsf{S} R'_{ji,k}
  \mathsf{S})$.  It is easy to prove that $\{ \overline{R}_{ij,k} \}$
  satisfies the corresponding normalization (Eq. (\ref{eq:normclon}) for cloning or  Eq. (\ref{eq:normlearn}) for learning)
 and that it gives the same value of $F$ as $R'_{ij,k}$.
  \qed
\end{Proof}

Eq. (\ref{eq:obs12permsymmetry}) together with the decomposition
(\ref{eq:obsdecompor21}) gives for $\forall (ab,c)\in \defset L$
\begin{align}\label{eq:obssimplifiedr}
  &\sigma_z r_{ab,c}^{\alpha} \sigma_z = r_{ba,c}^{\alpha} \quad
  r_{ab,c}^{\beta} = r_{ba,c}^{\beta} \quad
  r_{ab,c}^{\gamma}=r_{ba,c}^{\gamma}
\end{align}
where $\sigma_z = \begin{pmatrix}1 & 0 \\0& -1\end{pmatrix}$.

As a consequence of Eq. (\ref{eq:irrdecomp}) %and  Eq. (\ref{eq:obs12permsymmetry})
, the figure of
merit in Eq.~\eqref{eq:finfigm} can be written as
\begin{align}\label{eq:obsclonfigmerfinall}
  F =& \frac{1}{d^2}  \sum_{(ab,c)\in \defset L}  n(ab,c) \; \Tr [\ket{ijk}\bra{ijk}\sum_\nu  P^\nu\otimes r^\nu_{ab,c}] \nonumber\\
  = &   \sum_\nu \;\frac{1}{d}\sum_{(ab,c)\in \defset L}\Tr[\Delta_{ab,c}^\nu
  s_{ab,c}^\nu] \\
  =& F_\alpha + F_\beta + F_\gamma \nonumber
%   &F_\nu \equiv \sum_{(a,bc)\in \defset L}\Tr[\Delta_{a,bc}^\nu s_{a,bc}^\nu] \quad \nu\in\{\alpha,\beta\gamma\}
\end{align}
where
\begin{eqnarray}
  \Delta_{ab,c}^\nu &:=& \Tr_{\hilb{H}_\nu}[\ketbra{ijk}{ijk}], \\
  s_{ab,c}^{\nu} &:=& \frac{n(ab,c)}{d} r_{ab,c}^{\nu}
  \label{eq:obs12defnizdirridottiss}
\end{eqnarray}
and $ij,k$ is any triple of indices in the class denoted by $ab,c$.
Notice that
$n(xx,x)=d$, $n(xx,y)=n(xy,x)=n(xy,y)=d(d-1)$,
$n(xy,z)=d(d-1)(d-2)$ and
in the case $d=2$ $F_\gamma=0$ (i.e. does not appear).

 In particular, by direct
calculation we have
\begin{align} \label{eq:obsdeltas}
  &\Delta_{xx,x}^\alpha=
  \begin{pmatrix}
    \frac{2}{d+1}& 0 \nonumber\\
    0 & 0
  \end{pmatrix}, \quad\Delta_{xx,y}^\alpha= \frac12
  \begin{pmatrix}
    \frac{1}{d+1} & \frac{1}{\sqrt{d^2-1}} \nonumber\\
    \frac{1}{\sqrt{d^2-1}} & \frac{1}{d-1}
  \end{pmatrix},\nonumber\\
  &\Delta_{xy,y}^\alpha= \Delta_{xy,z}^\alpha= 0,\quad \Delta_{xy,x}^\alpha=\sigma_z \Delta_{xx,y}^\alpha \sigma_z  \nonumber\\
  &\Delta_{xx,x}^\beta = \frac{d-1}{d+1},\quad \Delta_{xx,y}^\beta = \Delta_{xy,x}^\beta =
  \frac{d}{2(d+1)},\nonumber\\
  & \Delta_{xy,y}^\beta = 1,\quad \Delta_{xy,z}^\beta = \frac12, \nonumber\\
  &\Delta_{xx,x}^\gamma =   \Delta_{xy,y}^\gamma =  0, \quad
  \Delta_{xx,y}^\gamma = \Delta_{xx,y}^\gamma =\frac{d-2}{2(d-1)},\nonumber \\
  &\Delta_{xy,z}^\gamma = \frac12.
\end{align}

\section{Optimal cloning}\label{sec:clon}

In this section we turn our attention to the cloning scenario.
Cloning is less restrictive than learning, since we allow the two states to be measured to be available at the same time as the single
use of the measurement device. The normalization condition for the $1
\rightarrow 2$ cloning %scenario is different from the $1 \rightarrow 2$ learning and
 reads
\begin{align}
  \sum_{ij,k} \ketbra{k}{k}_\lD \otimes R'_{ij,k} = I_\lD \otimes S_{\lA\lB\lC}
  \qquad \Tr_\lC[S] = I_{\lA\lB}
\end{align}
which implies the following
\begin{align}
  I_{\lA\lB}=&\Tr_\lC[R_{xx,x}]+(d-1)(d-2)\Tr_\lC[R_{xy,z}]\nonumber\\
  &+(d-1)\Tr_\lC[R_{xx,y}+R_{xy,x}+R_{xy,y}].
  \label{eq:normR}
\end{align}
From the commutation $[R_{ab,c},U^*_\lA\otimes U^*_\lB\otimes U_\lC]$ it
follows that $[\Tr_\lC[R_{ab,c}], U^*_\lA\otimes U^*_\lB]=0$
 and taking the
decomposition $R_{ab,c} = \sum_\nu P^\nu \otimes r_{ab,c}^\nu$ along with definition (\ref{eq:obs12defnizdirridottiss}),
the normalization constraint (\ref{eq:normR}) becomes
\begin{align}
  P^\pm=& P^\pm\sum_\nu\sum_{(ab,c)\in \defset L}\Tr_\lC [P^\nu \otimes s_{ab,c}^\nu]P^\pm .
\end{align}
%where $\delta_+=\beta$ and $\delta_-=\gamma$. Exploiting the
%relabeling symmetry (\ref{eq:obsperminvproperty}) and the permutation
%symmetry (\ref{eq:obs12permsymmetry}) we have
We take the trace of the previous equation to obtain the following equivalent formulation of the normalization constraints
\begin{align}
  d_+ &= d_\alpha \sum_{(ab,c)\in \defset L}s_{ab,c}^{\alpha,+,+} +
  d_{\beta}\sum_{(ab,c)\in \defset L} s_{ab,c}^{\beta},\label{eq:obsclonnormconst4} \\
  d_- &= d_\alpha \sum_{(ab,c)\in \defset L} s_{ab,c}^{\alpha,-,-} +
  d_{\gamma} \sum_{(ab,c)\in \defset L} s_{ab,c}^{\gamma}, \label{eq:obsclonnormconst3}
\end{align}
where $d_\pm\equiv \Tr[P^\pm]$, $d_\nu\equiv \Tr[P^\nu]$.
If we  introduce the notation
\begin{align}
  & s_{a,bc}^{\beta} :=
  \begin{pmatrix}
    s_{a,bc}^{\beta}&0\\
    0&0
  \end{pmatrix}
\qquad
  s_{a,bc}^{\gamma} :=
  \begin{pmatrix}
    0&0\\
    0&s_{a,bc}^{\gamma}
  \end{pmatrix}
 \nonumber \\
& \Pi^+ =
  \begin{pmatrix}
    1&0\\
    0&0
  \end{pmatrix}
  \qquad \Pi^- =
  \begin{pmatrix}
    0&0\\
    0&1
  \end{pmatrix}
\end{align}
the normalization constraints (\ref{eq:obsclonnormconst4}) and (\ref{eq:obsclonnormconst3})
can be rewritten as
\begin{align}\label{eq:obsclonfinalconst}
  &\Pi^+ \left( \sum_{\nu,(a,bc)\in \defset L} d_\nu s_{a,bc}^{\nu}
  \right) \Pi^+ = 
   \begin{pmatrix}
    d_+&0\\
    0& 0
  \end{pmatrix},
\nonumber\\
  &\Pi^- \left( \sum_{\nu,(a,bc)\in \defset L} d_\nu s_{(a,bc)}^{\nu}
  \right) \Pi^- = 
   \begin{pmatrix}
    0&0\\
    0& d_-
  \end{pmatrix}.
\end{align}
In order to solve the optimization problem we have to find the set ${
  \bf \mathsf{s} } := \{ s_{\ell}^{\nu}, \ell \in \defset{L}, \nu \in
\{\alpha, \beta \gamma \} \}$, $s_{\ell}^{\nu} \in
\mathcal{L}(\mathbb{C}^2), s_{\ell}^{\nu} \geq 0$ subjected to the
constraint (\ref{eq:obsclonfinalconst}) that maximizes the figure of
merit (\ref{eq:obsclonfigmerfinall}); we will denote as ${ \bf
  \mathsf{M} }$ the set of all the ${ \bf \mathsf{s} }$ satisfying Eq.
(\ref{eq:obsclonfinalconst}).  Since the figure of merit
(\ref{eq:obsclonfigmerfinall}) is linear and the set ${ \bf \mathsf{M}
}$ is convex, a trivial result of convex analysis states that the
maximum of a convex function over a convex set is achieved at an
extremal point of the convex set.  We now give two necessary
conditions for a given ${ \bf \mathsf{s} }$ to be an extremal point of
${ \bf \mathsf{M} }$.  Let us start with the following
\begin{definition}[Perturbation]
  Let ${ \bf \mathsf{s} }$ be an element of ${ \bf \mathsf{M} }$.  A
  set of hermitian operators ${ \bf \mathsf{z} }:= \{ z_{\ell}^{\nu}
  \}$ is a \emph{perturbation} of ${ \bf \mathsf{s} }$ if there exists
  $\epsilon \geq 0$ such that
  \begin{align}
    { \bf \mathsf{s} } + h { \bf \mathsf{z} } \in { \bf \mathsf{M} }
    \qquad\forall h \in [-\epsilon, \epsilon ]
  \end{align}
  where we defined ${ \bf \mathsf{s} } + h { \bf \mathsf{z} } := \{
  s_{\ell}^{\nu} + h z_{\ell}^{\nu}|h\in[-\epsilon,\epsilon]\}$.
\end{definition}
By the definition of perturbation it is easy to prove that an element
${ \bf \mathsf{s} }$ of ${ \bf \mathsf{M} }$ is extremal if and only
if it admits only the trivial perturbation $z_{\ell}^{\nu}=0$ $\forall
\ell, \nu $.  We now exploit this definition to prove two necessary
conditions for extremality.
\begin{lemma}\label{lem:secondextremalcond}
Let  ${ \bf \mathsf{s} }$ be an extremal element of ${ \bf \mathsf{M} }$.
Then $s_{\ell}^{\nu}$ has to be rank one for all $\ell, \nu$.
\end{lemma}
\begin{Proof}
  Suppose that there is a $s_{\ell '}^{\nu '} = \left(
   \begin{array}{cc}
     a & b \\
     c & d
   \end{array}
\right)
 \in { \bf \mathsf{s} }$
which is not rank one;
then there exist $\epsilon$ such that
${ \bf \mathsf{z} } := \{0,\dots,0,z_{\ell '}^{\nu '},0, \dots, 0  \}$,
$ z_{\ell '}^{\nu '} =
\left(
   \begin{array}{cc}
     0 & 1 \\
     1 & 0
   \end{array}
\right)$ is an admissible perturbation. \qed
\end{Proof}

The above lemma tells us that without lost of generality we can assume the optimal ${ \bf
  \mathsf{s} }$ to be a set of rank one matrices.  Let us now consider
a set ${ \bf \mathsf{s} }$ such that $s_{\ell}^{\nu}$ is rank one for
all $\ell, \nu$; any admissible perturbation ${ \bf \mathsf{z} }$ of
${ \bf \mathsf{s} }$ must satisfy
\begin{align}
& z_{\ell}^{\nu} = c_\ell^\nu s_{\ell}^{\nu} \qquad c_\ell^\nu \in  \mathbb{R} \label{eq:perturbprespos}\\
&
  \Pi^+
\left(
 \sum_{\nu,\ell} d_\nu
c_\ell^\nu s_{\ell}^{\nu}
\right)
\Pi^+ =
 \Pi^-
 \left(
\sum_\nu
 d_\nu
c_\ell^\nu s_{\ell}^{\nu}
 \right)
\Pi^- = 0.
\label{eq:perturbpreservnorm}
\end{align}
where the constraint (\ref{eq:perturbprespos})
is required in order to
have $s_{\ell}^{\nu} + h  z_{\ell}^{\nu} \geq 0$,
 while Eq.  (\ref{eq:perturbpreservnorm})
tells us that
${ \bf \mathsf{s} } + h { \bf \mathsf{z} }$ satisfies
the normalization (\ref{eq:obsclonfinalconst}).
Let us now consider the map
\begin{align}
&f: \mathcal{L}(\mathbb{C}^2) \longrightarrow \mathbb{C}^2 \qquad
f(A) :=
\left(
 \begin{array}{c}
\Tr[\Pi^+   A]\\
\Tr[\Pi^-  A]
 \end{array}
\right)
 \nonumber \\
&f
\left(
 \begin{array}{cc}
 a&b\\
 c&d
 \end{array}
\right)
=
\left(
 \begin{array}{c}
a\\
d
 \end{array}
\right) \nonumber
  \end{align}
exploiting this definition
Eq. (\ref{eq:perturbpreservnorm})
becomes
\begin{align}
 \sum_{\nu, \ell} c_\ell^\nu f(s^\nu_\ell) =
\left(
 \begin{array}{c}
   0 \\
   0
 \end{array}
\right).
\end{align}
Suppose now that
the set $ \overline{{ \bf \mathsf{s}} }$
has $N \geq 3$ non-zero elements;
then $\{ f(\overline{s}_\ell^\nu) \}$ is a set
of $N \geq 3$ vectors of $\mathbb{C}^2$
that cannot be linearly independent.
That being so, there exists a set of
coefficients $\{ c_\ell^\nu \}$ such that
$ \sum_{\nu, \ell} c_\ell^\nu f(s^\nu_\ell) =0$
and then
$ z_{\ell}^{\nu} = c_\ell^\nu \overline{s}_{\ell}^{\nu}$
is a perturbation of $ \overline{{ \bf \mathsf{s}} }$.
We have then proved the following lemma.
\begin{lemma}\label{lem:seconextremalitycond}
  Let $ { \bf \mathsf{s}} $
be an extremal element of $\defset{M}$.
Then  $ { \bf \mathsf{s}} $ cannot have
more than $2$ non-zero elements.
\end{lemma}
Lemma \ref{lem:secondextremalcond} and Lemma
\ref{lem:seconextremalitycond} provide two necessary %sufficient
conditions for extremality that allow us to restrict the search
of the optimal $ { \bf \mathsf{s}} $ among the ones that satisfy
\begin{align}
{ \bf \mathsf{s}} = \{s_{\ell'}^{\nu'}, s_{\ell''}^{\nu''}  \}
\qquad \rank(s_{\ell'}^{\nu'})=\rank(s_{\ell''}^{\nu''})=1 \nonumber\\
 \Pi^i
\left(
 \sum_{\nu,\ell} d_\nu
s_{\ell}^{\nu}
\right)
\Pi^i =
 d_i \quad  i = +,-
\end{align}
The set of the above $ { \bf \mathsf{s}} $
is small enough to allow us to  compute
the value of $F$ for all the possible cases.
It turns out that there are two choices achieving the highest value of fidelity
\begin{align}
F = \frac{4}{3d}. \label{eq:clonfinalf}
\end{align}
They are defined by
${  \bf \mathsf{s}} = \{    s^\alpha_{xx,x},  s^\alpha_{xy,x} \}$ and ${  \bf \mathsf{s}} = \{    s^\alpha_{xx,x},  s^\alpha_{xy,y} \}$, where

\begin{align}
  &  s^\alpha_{xx,x} =
\left(
 \begin{array}{cc}
\frac{9d_+ -1}{9d} &0\\
0&0
 \end{array}
\right)\equiv A,
\qquad
   & B \equiv
\left(
 \begin{array}{cc}
\frac{1}{9d}&\frac{\sqrt{d_-}}{3d}\\
\frac{\sqrt{d_-}}{3d}&\frac{d_-}{d}
 \end{array}
\right), \nonumber \\
& s^\alpha_{xy,x} = B,   \quad s^\alpha_{xy,y} = \sigma_z B \sigma_z.
\end{align}
From the linearity of the link product and our figure of merit it follows that also any convex combination of the above two strategies will give the optimal performance. In the rest of the paper we consider the equal convex combination of the above two strategies
\begin{align}
  &  s^\alpha_{xx,x} =A,\qquad  s^\alpha_{xy,x} = \frac12 B,   \quad s^\alpha_{xy,y} = \frac12 \sigma_z B \sigma_z,
  \label{finsol}
\end{align}
because it treats the two clones in the same way.
%\subsection{Form of replicated measurement}
Using Eq. (\ref{eq:reppovm}) one can derive the form of the replicated POVM corresponding to the above choice of the optimal generalized instrument.
\begin{eqnarray}
G_{ii}&=&\left( 1- \frac{2}{9d(d+1)} \right) P^+ (E^{(U)}_i \otimes I_\lB) P^+ \nonumber \\
G_{ij}&=&\frac{1}{d-1}[Q^+ (E^{(U)}_i \otimes I_\lB) Q^+ + Q^- (E^{(U)}_j \otimes I_\lB) Q^- ], \nonumber
\end{eqnarray}
where $Q^\pm=1/\sqrt{9d(d+1)}\; P^+ \pm 1/\sqrt{2}\; P^-$.

\subsection{Realization scheme for the optimal cloning Network}
\label{sec:realization}
In this section we describe the inner structure
of the optimal cloning network.
First we notice that the choice from Eq. (\ref{finsol})
corresponds to the generalized
instrument
\begin{align}
 &R_{ii}   =
% \ketbra{i}{i} \otimes r^{\alpha}_{ii,i} =
\ketbra{i}{i}
\otimes
\frac{9d_+   -1}{9d}
\sum_k \ketbra {\Psi^+_k}{\Psi^+_k}
\nonumber
 \\
& R_{ij}   =
% \ketbra{i}{i} \otimes r^{\alpha}_{ij,i}
% +
% \ketbra{j}{j} \otimes r^{\alpha}_{ij,j}
%  =
\ketbra{i}{i}
\otimes
\frac1{2(d-1)}
\sum_k \ketbra {\phi_k}{\phi_k} +
\nonumber\\
&+
\ketbra{j}{j}
\otimes
\frac1{2(d-1)}
\sum_k  \tilde{\sigma}_z \ketbra {\phi_k}{\phi_k} \tilde{\sigma}_z
\label{newoptimalnet}
\\
&\ket{\phi_k} = \sqrt{\frac1{9d}}\ket{\Psi^+_k} +
\sqrt{\frac{d_-}{d}}\ket{\Psi^-_k}
\nonumber \\
&\tilde{\sigma}_z \ket{\Psi^\pm_k} = \pm  \ket{\Psi^\pm_k}.
\nonumber
\end{align}
% which achieves the maximum value of the fidelity
%derived in Eq. \eqref{eq:obsclonoptimalfide}.

The generalized instrument $\vec R$ can be realized by the
following network
\begin{align}
   \begin{aligned} \Qcircuit @C=0.8em @R=1.5em
{
&  \ustick{\lA}  \qw & \multigate{1}{SWAP} & \ustick{\lC} \qw  &
  \measureD{{\vec E}^{(U)}} & \ustick{\lD}    \cw&  \multipuregate{2}{f} \cw &&&\\
& \ustick{\lB} \qw & \ghost{SWAP} &  \ustick{K} \qw &
 \measureD{\vec I} &  \pureghost{f} \\
& \lstick{\ket{+}}& \ctrl{-1}&   \ustick{L} \qw
& \measureD{\vec P} &  \cw & \pureghost{f} \cw &&&&
}
  \end{aligned}
\end{align}
The first step consists of a control SWAP gate, which is described by the
unitary
\begin{align}
U_{CS}& =  T_{\lA \to \lC} \otimes T_{\lB \to \lK} \otimes \ketbra{0}{0}_\lL + \nonumber \\
& + T_{\lA \to \lK} \otimes T_{\lB \to \lC} \otimes \ketbra{1}{1}_\lL \nonumber
\end{align}
 with the control qubit prepared in the state $\ket{+} =
\frac{1}{\sqrt{2}}(\ket{0} + \ket{1} )$.  We defined $T_{X \to Y}=\sum_i \ket{i}_Y\bra{i}_X$ and 
named $\hilb{H}_\lL$ the $2$-dimensonal Hilbert space
of the control qubit with $\{  \ket{0}, \ket{1} \}$ being
 an orthonormal basis on $\hilb{H}_\lL$.

In the second step
we have three commuting actions:
\begin{itemize}
\item  the single use of the measurement device
${\vec E}^{(U)}$ is applied on system $\lC$ and its outcome is recorded
on a classical memory $\lD$
\item system $\lK$ is discarded
\item system $\lL$ undergoes a $3$-outcome measurement
described by the POVM
$\vec P$
defined as follows:
\begin{align}
 &P_1 = \frac{(9d(d+1) -2)}{9d(d+1)} \ketbra{+}{+} \nonumber \\
 &  P_2 = \ketbra{\psi}{\psi} \quad
 P_3 = \sigma_z \ketbra{\psi}{\psi} \sigma_z
\nonumber \\
&\ket{\psi} =
 \sqrt{\frac{1}{9d(d+1)}} \ket{+} +
\sqrt{\frac{1}{2}} \ket{-}
\end{align}
\end{itemize}

The last step is  just a classical processing $f$
of the outcome $k$ of the measurement % $\{E_k^{U}\}$
${\vec E}^{(U)}$ and of the outcome $n$ of POVM $\vec P$. %$\{ P_n \}$.
The function $f$ that produces the  outcome $(i,j) = f(k,n)$
of the network is defined as follows:
\begin{align}
f(k,n) =
\left \{
\begin{array}{ccl}
(k,k) & & \mbox{if } n=1 \\
(k,j) & j \neq k & \mbox{if } n=2 \\
(j,k) & j \neq k  &\mbox{if } n=3
\end{array}
\right. ,
\end{align}
where the outcome $j$ in the second and third case
is randomly generated with flat distribution.

In order to prove that  this network
is described by the generalized instrument in Eq~\eqref{newoptimalnet}
we first realize that  the action of the POVM $\vec P$ and
of the processing $f$  can be represented by the bipartite POVM
$\vec Q$ on systems $\lD$ and $\lL$ defined as
\begin{align}
 &Q_{i,j} =
\left \{
\begin{array}{ll}
\ketbra{i}{i} \otimes \frac{(9d(d+1) -2)}{9d(d+1)} \ketbra{+}{+}
&
\mbox{if } i=j \\
&\\
\ketbra{i}{i} \otimes \frac{\ketbra{\psi}{\psi}}{d-1}
+
\ketbra{j}{j} \otimes
\frac{\sigma_z \ketbra{\psi}{\psi} \sigma_z}{d-1}
&
\mbox{if } i \neq j \\
\end{array}
\right.
 \nonumber  \\
 \nonumber  \\
&   \begin{aligned} \Qcircuit @C=0.8em @R=1.5em
{
 &\ustick{\lC} \qw  &
  \measureD{{\vec E}^{(U)}} & \ustick{\lD}    \cw&
 \pureghost{\vec Q} \cw &&&\\
   &
 && &
 \pureghost{\vec Q}\\
  &\qw&
\qw& \qw&
\multimeasureD{-2}{\vec Q}
}
  \end{aligned}
 \!\!\!\!\!\!\!\!\! = \;\;\;
   \begin{aligned} \Qcircuit @C=0.8em @R=1.5em
{
&\ustick{\lC} \qw  &
  \measureD{{\vec E}^{(U)}} & \ustick{\lD}    \cw&\multipuregate{2}{f} \cw&&&\\
 &  &
 &  \pureghost{f}\\
 &  \ustick{\lL} \qw
& \measureD{\vec P} &  \cw  & \pureghost{f}  \cw&&&
}
  \end{aligned}
\end{align}

Finally, one can check the identity
\begin{align}
 R_{ij} = \ketbra{+}{+} * |{U_{CS}} \>\!\> \< \!\<{U_{CS}}| * (Q_{i,j}\otimes I_\lK)
\end{align}

It is worth to notice that the optimal cloning of measurement device
has some features in common with the optimal cloning of unitaries.
Both in the cloning of unitaries and in the cloning of
von Neumann measurements
the first step is to perform a control-SWAP of the two input systems
with the control qubit prepared in the superposition
$\frac{1}{\sqrt{2}}(\ket{0} + \ket{1} )$.
We could give an intuitive explanation of this feature in terms of
quantum parallelism:
for a bipartite input $\ket{\chi}_{0}\ket{\xi}_1$
the unknown measurement %example is run on both sides through the
acts on both input states via a superposition
$\sqrt{\frac{1}{2}} (\ket{\chi}_{2}\ket{\xi}_A
+\ket{\xi}_{2} \ket{\chi}_A  ) $.

\section{Optimal learning}\label{sec:learning}

Our goal in this scenario is to create two replicas of the measurement after it was used once.
Let us consider the normalization constraint for the generalized
instrument $R_{ij}$. Since $\sum_{i,j}R_{ij}$ has to be a
deterministic network, we have
\begin{align}
  \sum_{ijk} \ket{k}\bra{k}_\lD \otimes R'_{ij,k}= I_{\lA\lB\lD}\otimes \rho_\lC,\quad \Tr[\rho]=1 \label{eq:obsnormL12}
\end{align}
where $\rho$ has to be positive operator. The commutation relation
(\ref{eq:obscommutr}) implies $[\rho,U]=0$ and so we have
$\rho=\frac{1}{d}I_\lC$.  Writing $I_{\lA\lB\lC\lD}$ as $\sum_k \ket{k}\bra{k}_\lD
\otimes ( I_{m_\alpha}\otimes P^{\alpha}+P^{\beta}+P^{\gamma})$ we can
rewrite the normalization conditions as follows
\begin{align}
  &\sum_{(ab,c)\in \defset L}s_{ab,c}^{\nu}=\frac{1}{d}
  \quad
  \nu = \beta, \gamma \nonumber\\
  &\sum_{(ab,c)\in \defset L}s_{ab,c}^{\alpha}=\frac{1}{d}I_{m_\alpha}
\label{eq:obsnormL21final}
\end{align}
Let us now maximize the figure of merit under these constraints.
The maximization of $F_\beta$ and $F_\gamma$
is  simple and yields
\begin{align}
&  F_\beta = \frac1{d^2} \qquad
 F_\gamma = \frac1{2d^2}\\
 &s_{xx,x}^{\beta}=s_{xy,x}^{\beta}=s_{xy,y}^{\beta}=s_{xy,z}^{\beta}=0 \nonumber \\
 &s_{xx,x}^{\gamma}=s_{xx,y}^{\gamma}=s_{xy,x}^{\gamma}=s_{xy,y}^{\gamma}=0, \nonumber \\
 &s_{xx,y}^{\beta}=s_{xy,z}^{\gamma}=\frac{1}{d}. \nonumber \\
\end{align}
Let us now consider the maximization of $F_\alpha$;
the normalization constraint for the $\alpha$ subspace gives
\begin{align}
  &\sum_{(ab,c)\in \defset L} s_{ab,c}^{\alpha, +, +}=\frac{1}{d},
  &\sum_{(ab,c)\in \defset L} s_{ab,c}^{\alpha, +, -}=0, \nonumber \\
  &\sum_{(ab,c)\in \defset L} s_{ab,c}^{\alpha, -, -}=\frac{1}{d},
  &\sum_{(ab,c)\in \defset L} s_{ab,c}^{\alpha, -, +}=0.
  \label{eq:obs12constraintperalpha}
\end{align}
Inserting the explicit expression of the $\Delta_{ab,c}^\alpha$ into
Eq. (\ref{eq:obsclonfigmerfinall}) and taking into account Eq.(\ref{eq:obssimplifiedr}) we have

\begin{align}
 d F_\alpha =& \Tr \left[\begin{pmatrix} s_{xx,x}^{\alpha, +, +} & s_{xx,x}^{\alpha, +, -}\\
  s_{xx,x}^{\alpha, -, +} & s_{xx,x}^{\alpha, -, -}
\end{pmatrix}
\begin{pmatrix}
  \frac2{d+1}&0\\
  0&0
\end{pmatrix}\right]+\nonumber\\
&\Tr\left[\begin{pmatrix}
  s_{xy,x}^{\alpha, +, +} & s_{xy,x}^{\alpha, +, -}\\
  s_{xy,x}^{\alpha, -, +} & s_{xy,x}^{\alpha, -, -}
\end{pmatrix}
\begin{pmatrix}
  \frac1{d+1}&\frac1{\sqrt{d^2-1}}\\
  \frac1{\sqrt{d^2-1}}&\frac1{d-1}
\end{pmatrix}
\right] = \nonumber \\
&= \frac{2s_{xx,x}^{\alpha, +, +}}{d+1}+
\frac{s_{xy,x}^{\alpha, +, +}}{d+1}+\frac{s_{xy,x}^{\alpha, -, -}}{d-1}+\frac{2s_{xy,x}^{\alpha, +, -}}{\sqrt{d^2-1}} \leq \nonumber \\
% &\leq\frac2{(d+1)}\left( \frac1d - 2s_{xy,x}^{\alpha, +, +}  \right)+
% \frac{s_{xy,x}^{\alpha, +, +}}{d+1}+\frac{s_{xy,x}^{\alpha, -, -}}{d-1}+
% 2\sqrt{\frac{s_{xy,x}^{\alpha, +, +}s_{xy,x}^{\alpha, -, -}}{d^2-1}}\leq \nonumber \\
& \leq
\frac{5d-3}{2d(d^2-1)}
-\frac{3s_{xy,x}^{\alpha, +, +}}{d+1}+
2\sqrt{\frac{s_{xy,x}^{\alpha, +, +}}{2d(d^2-1)}}
\label{eq:obs12upperboundFa}
\end{align}
where in the derivation of the bound (\ref{eq:obs12upperboundFa}) we
used the positivity of $s_{xy,x}^{\alpha}$ and the constraints
(\ref{eq:obs12constraintperalpha}).  The upper bound
(\ref{eq:obs12upperboundFa}) can be achieved by taking
\begin{align}
  &s_{xx,x}^\alpha=
  \begin{pmatrix}
    \frac1{d} - 2a &0\\
    0&0
  \end{pmatrix}\quad
  s_{xy,x}^\alpha=
  \begin{pmatrix}
    a&\sqrt{ \frac{1}{2d} a}\\
    \sqrt{ \frac{1}{2d} a}&\frac{1}{2d}
  \end{pmatrix}, \nonumber\\
  & s_{xy,z}^\alpha=s_{xx,y}^\alpha=0
\end{align}
where we defined $a := s_{xy,x}^{\alpha,+,+}$.
Eq. (\ref{eq:obs12upperboundFa}) gives the value of $F_\alpha$
as a function of $a$; the maximization of $F_\alpha(a)$ with the constraint
$0\leq a \leq \frac1{2d} $ is easy and gives
\begin{align}
  F_\alpha = \frac{4(2d-1)}{3d^2(d^2-1)} \qquad \mbox{for }  a = \frac{d+1}{18d(d-1)} .
\end{align}
and then for $d \geq 3$ we have
\begin{align}\label{eq:maxfidelear12}
  F = F_\alpha + F_\beta + F_\gamma =
  \frac{9d^2+16d-17}{6d^2(d^2-1)}\sim\frac3{2d^2} .
\end{align}
For $d=2$ the invariant subspace $\hilb{H}_\gamma$
does not appear and the fidelity becomes $F = F_\alpha +F_\beta = \frac{7}{12}$.

Using Eq. (\ref{eq:reppovm}) it is possible to derive the form of the replicated POVM corresponding to the optimal generalized instrument.
\begin{eqnarray}
G_{ii}&=&\frac{16d-2}{9d(d^2-1)} P^+ (E^{(U)}_i \otimes I_\lB) P^+  +\frac{d^2-3}{d(d^2-1)}P^+ \nonumber \\
G_{ij}&=&\frac{1}{d-1}[Q'^+ (E^{(U)}_i \otimes I_\lB) Q'^+ + Q'^- (E^{(U)}_j \otimes I_\lB) Q'^- ] \nonumber \\
& &+ \frac{2}{d(d-1)^2(d-2)}P^- (E^{(U)}_i \otimes I_\lB + E^{(U)}_j \otimes I_\lB) P^-  \nonumber \\
& &+\frac{d-3}{(d-1)^2(d-2)}P^-  \nonumber
\end{eqnarray}
where $Q'^\pm=1/\sqrt{9d(d-1)}\; (P^+ \pm 3\; P^-)$.

\begin{figure}[h]
    \includegraphics[width=8.5cm ]{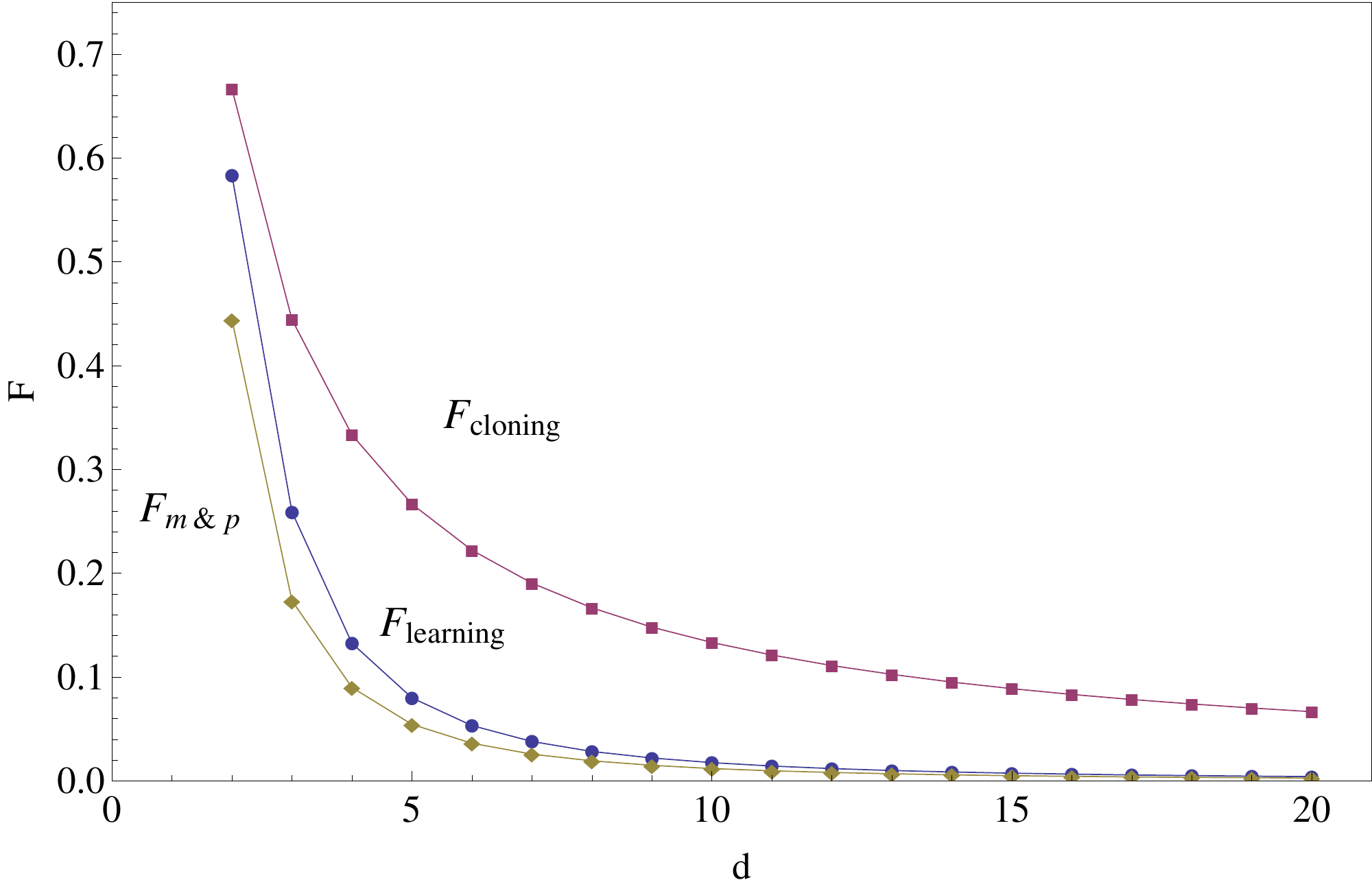}
    \caption{\label{fig:comparison} Optimal $1\rightarrow 2$ cloning and learning of a measurement
      device: we present the values of $F$ for different values of
      the dimension $d$.  The squared dots represent the optimal
      $1\rightarrow 2$ cloning, the round
      dots represent the optimal $1\rightarrow 2$ learning and lowest curve corresponds to a strategy in which one performs the optimal estimation followed by the preparation of the estimated measurement.  }
\end{figure}

One can now compare the performance of the optimal $1\rightarrow 2$ cloning and learning. The optimal values of $F$ depending on the dimension $d$ are plotted on Figure \ref{fig:comparison}. 
As expected the optimal cloning strategy largely outperforms the optimal learning strategy with a fidelity, which is a factor $d$ larger, as one can see from  Eqs. (\ref{eq:clonfinalf}) and (\ref{eq:maxfidelear12}). Similar distinction arises also for comparison of cloning and learning of unitary channels (for details see refs \cite{learnunit}).
It is also worth noting that the optimal learning strategy achieves a
greater fidelity than the incoherent strategy in which one first make
the optimal estimation of the measurement and then conditionally
prepares two copy of the estimated measurement (one can prove that for
this last strategy one has $F_{m \& p} = (\frac{d+2}{d(d+1)})^2$ \cite{nota}).

% stress that the difference between learning and cloning is the causal order.

\section{Conclusions}\label{conc}
In the present paper we focused on
$1\rightarrow 2$ cloning and $1\rightarrow 2$ learning of von Neumann measurements. Even though both problems can be easily formulated in the usual language of quantum mechanics, the necessity to handle the measurement outcome in the remaining part of the scheme makes the optimization complicated and requires suitable mathematical tools. We represented the unknown measurement to be replicated as a measure\&prepare channel and we employed framework of quantum combs to perform the network optimization. Thanks to symmetries of the figure of merit the problem was simplified and solved for arbitrary dimension of the measurement's Hilbert space $d$. In section (\ref{sec:realization}) we proposed a realization of optimal $1\rightarrow 2$ cloning of measurements. The proposed scheme has some similarities to optimal cloning of unitary transformations, since they both begin by the control-swap operation, which reflects the presence of quantum parallelism.
In this paper we expolit the measure\&prepare representation  of von Neumann measurement that allowed us to deal with feed forward of classical information in quantum networks. These tool could be in principle used to tackle other quantum information processing tasks
 in which classical information is involved e.g. estimation and cloning of quantum instruments.

\section*{Acknowledgments}
This work has been supported by the European Union through FP$7$ STREP project COQUIT and by the Italian Ministry of Education through grant PRIN 2008
Quantum Circuit Architecture.


\begin{thebibliography}{99}
\bibitem{nota}
We omit the derivation of this expression for sake of
simplicity. However it can be derived exploiting the same techniques
as for the derivation of the optimal learning in Sec. \ref{sec:learning}.
\end{thebibliography}
\end{document}